# Women's inheritance rights reforms and impact on women's empowerment: evidence from India

**Minali Grover**

Indian Institute of Management Indore, India

**Ajay Sharma**

Indian Institute of Management Indore, India

Global Labor Organization, Germany

**Declaration of Interest Statement:** We know of no conflicts of interest associated with this publication, and there has been no significant financial support for this work that could have influenced its outcome. As the corresponding author, I confirm that the manuscript has been read and approved for submission by all the named authors.

**Data availability:** The data used in the study is openly available in Indian Human Development Survey (2004-05) site, https://ihds.umd.edu/.

**Corresponding Address:** Minali Grover, Indian Institute of Management Indore, Prabandh Shikhar, Rau-Pithampur Road, Indore, Madhya Pradesh, India-453556; e-mail: f21minalig@iimidr.ac.in

Ajay Sharma, J-206, Academic Block, Indian Institute of Management Indore, Prabandh Shikhar, Rau-Pithampur Road, Indore, Madhya Pradesh, India-453556; e-mail: ajays@iimidr.ac.in, ajaysharma87@gmail.com

*Abstract*

*This paper explores the influence of inheritance rights on women' empowerment in India. We employ the quasi-natural experiment framework wherein; five states amended the Hindu Succession Act (HSA) from 1976 to 1994 before it was federally amended in 2005. Further, we apply difference-in-difference (DID) strategy and consider triangulation approach to identify women empowerment indicators namely: access to resources, agency, and outcomes to measure varying dimensions of empowerment. Using the India Human Development Survey (IHDS-I), our results indicate a positive impact on marriage choice, intimate partner violence, physical, and civil autonomy. However, negative impact on household autonomy and no significant on economic participation for women exposed to state amendments. Further, exploring the heterogeneities in terms of socio-economic status, location, level of patriarchy in a state, gender of the head of the household. Overall, the study highlights that the impact of inheritance law is not unfirm across different groups.*

**Keywords:** *Inheritance law, Hindu Succession Act, India, Women Empowerment, Autonomy, Gender*

**JEL Classification:** *J16, J21, K11*

## 1. Introduction

Women's empowerment has been a salient issue in developing countries, especially in South Asia, where there exist restrictive cultural norms and a patriarchal society. Apart from having intrinsic importance for individuals, women's empowerment also contributes to economic enhancement, better children-related outcomes (Duflo, 2003; Thomas, 1992), human capital investment (Luke & Munshi, 2005), productivity (Anik & Rahman, 2021), and other aspects. However, definition of female empowerment is largely abstract and multi-faceted as it is a sum of both the individual choices and how these individual choices are accepted by the society at large (Anderson, 2022; 2024; Kabeer, 1999). Thus, any attempt to empower women goes through various stages stemming from bargaining power within the household, choice over marriage, freedom over mobility, decision-making over financial matters, control over economic resources, and so on. Moreover, these indicators might not move in tandem that is, greater household autonomy does not necessarily lead to greater empowerment in public sphere and vice versa[1] (Anderson, 2024). Past scholars (see, for example, Kabeer, 1999) also demonstrate that empowering women is not a one-step process or outcome rather a "process of change" spanning through the accessibility of resources (pre-conditions) that translates to gaining agency (control) towards decision-making and finally, leading to outcomes (achievements) such as participation in the labour force, equal gender roles, reduction in reports of injustice, domestic violence, and so on. In a similar context, Kishor (2000) defines pre-conditions for women empowerment by identifying sources of empowerment that aid in improving their control or decision-making such as, education, assets owned,

---

[1] For instance, in India, a third of the seats are reserved for women councils of the Panchayat Raj institutions since 1992 (more than 30 years ago). However, this amendment of increasing political participation at the public sphere had not led to changes in terms of economic participation of women where India stands at a dismay low figure of around 30 percent (Anderson, 2024). Moreover, there exist substantial evidence of how increase in female economic empowerment increases conflict within the household thereby, leading to rise in intimate partner violence (Luke & Munshi, 2011)

employment, control over earnings, and so on. Building on this view, providing property or inheritance rights through institutional changes that increases chances of land ownership can be a direct source of empowerment for women.

Institutional amendments along with development in the societal norms are primary two channels through which a radical change in women empowerment can be stemmed. However, institutional changes itself are contingent on social norms and development of the country. For instance, while developed countries such as the U.S.A and U.K. observed progress in economic rights followed by political rights and equal treatment in the labour, the trajectory for the developing countries is in contrast. The timeline of the developing countries begins with legal and political rights followed by changes in the economic rights and societal norms (Anderson, 2022; Doepke et al., 2012). There exist ample of evidence how extending electoral quotas for women in political institutions render positive long-term impact in altering stereotypes about female leadership and future opportunities for female leaders even when quotas are removed (Beaman et al., 2009; Bhavnani, 2009). Similarly, extending economic rights through gender equal inheritance rights render several favourable impacts. For instance, Allendorf (2007) shows women owning land has better intra-household decision-making power in Nepal while similar evidence is portrayed for the case of Egypt (Khodary, 2018), India (Mookerjee, 2019; Bose & Das, 2020), and Kenya (Harari, 2019) among others. For India, extant studies analysing policy changes in the inheritance rights found increase in labour supply, education, autonomy indicators for women (Heath & Tan, 2020; Mookerjee, 2019; Roy, 2015). Furthermore, another institutional change in prohibiting dowry under, Dowry Prohibition Rules (1985) led to decreasing dowries (Alfano, 2017; Bhat & Thakur, 2024), lifetime fertility of women (Alfano, 2017), and increase human capital investment (Calvi & Keskar, 2023). However, amendments at the institutional level and social norms might not move in conjunction always. In contrast, there might be a backlash effect from husbands or family members to gain control back from their wives or female members. For instance, Calvi & Keshar (2023) provide evidence for increase in domestic violence and decrease in women's bargaining power as a result of Dowry Prohibition Act. Another example stems from how state amendments in inheritance law have also led to increase in dowries (Roy, 2015), female child mortality (Rosenblum, 2015), and suicides (Anderson & Genicot, 2015). Overall, these studies indicate how progressive institutional changes could interact inversely with the restrictive gender norms thereby, leading to negative outcomes.

Overall, whether channels of increasing women empowerment or linkage within different indicators or spheres of women empowerment are considered, both are webbed into complex mechanism interacting household, societal, economic, legal, and political dynamics. As a result, when analysing women empowerment, it is important to uncover several dimensions and spheres rather than relying on one broad indicator. In this paper, following Kabeer's (1999) approach of defining women empowerment, we adopt a three-step process; access to resources, agency, and achievements to measure the potential impact of change in inheritance rights in India on women's empowerment using the India Human Development Survey (IHDS) (2004–05). Inheritance is part of both state and federal subjects in India. Therefore, five states were able to pass institutional amendments providing women with equal inheritance rights between 1976 to 1994 much before a federal amendment was made in Hindu Succession Amendment Act (HSAA) in 2005 which was applicable nation-wise. This provided a quasi-natural experiment to be explored for researchers.

Given, nearly 97 percent of property is jointly owned (Roy, 2015), and land-related disputes affect 3.1 million people (Brule, 2020), with more than half of the land and property disputes occurring within the family (Daksh, 2016); analysing its impact becomes imperative for India. As for women, who derive 75 percent of their income from agriculture, their land ownership stands at a dismayingly low figure of 13 percent (Brule, 2020). Thus, overall, highlighting the disparities in land ownership and the significance of HSAA for women. Moreover, in India where economic responsibilities are linked with deep-rooted social norms, ownership rights are not merely means of livelihood for people; they further translate into social responsibilities stretching over the life cycle including decisions related to children and parents and inter-generational wealth transfer. Thus, interacting with various household and economic dynamics. Since, in the Indian context, the household head (referred to as "karta") has the principal responsibility over the family's property, they are also the prime decision-makers. However, by encompassing rights to land inheritance for daughters, it reinforces gender-egalitarian relationships within the family. Theoretically, following the bargaining model framework, land inheritance could increase the outside option available for women; thereby, providing them with greater autonomy and mobility. Empirically, land inheritance in women's names provides them with greater resources to look after their families, increase in human capital investments, and a greater ability to overcome domestic violence (Allendorf, 2017; Khadoary, 2018; Panda & Agarwal, 2005). Since the direct impact of inheritance of joint property to daughters cannot be observed, studies have measured the potential impact for women exposed to reform through the state amendments passed in five states between 1976 to 1994 before the federal law was passed. Several past studies have measured the HSAA impact on development-related measures such as; education (Suteau, 2020), health outcomes (Heath & Tan, 2019), dowry payments (Roy, 2015), working status (Heath & Tan, 2019; Suteau, 2020), intergenerational outcome (Bose & Das, 2017; Tandel et al., 2023), age of marriage (Suteau, 2020), land inherited (Deininger et al, 2013; Roy, 2015), child-related outcomes (Bose & Das, 2020; Tandel et al., 2023), suicides (Anderson & Genicot, 2015), violence (Amaral, 2017; Bose & Das, 2020), and so on.

We summarize the extant literature in three broad areas. First, studies analyzing the first-level effect of inheritance law. There is mixed evidence for the direct impact of inheritance law on the likelihood of receiving inheritance (Roy, 2015; Deininger et al., 2013). On one hand, Deininger et al., (2013) finds higher likelihood for women to inherit land whereas Roy (2015) founds no significant impact[2]. However, there are compensating mechanisms through which wealth is transferred to daughters as a result of HSAA. For instance, women exposed to the reform and belonging to the age group 11-15 receive 50 percent more dowry payments whereas girls younger than 10 years received 1.2-1.7 years of higher education (Roy, 2015)[3]. Second, researchers further explore the second-level effect involving changes in the decision-making within and outside the household as a result of the egalitarian inheritance norms. With increased education, women autonomy outcomes related to household decision-making and mobility also improved (Roy, 2008; Heath & Tan, 2019; Mookerjee, 2019). The results indicate that HSAA increased the women's autonomy by 0.17 standard deviations (Heath & Tan, 2019) whereas Mookerjee (2019) shows a rise by 2-5 percent for different autonomy indicators. However, despite an increase in the bargaining power and human capital (Deininger, 2019; Roy, 2015), there exists mixed evidence for an increase in the labour force participation for women (Heath & Tan, 2019; Suteau, 2020) or changes in domestic violence (Amaral, 2017; Bose & Das, 2020). Lastly, the next channel of transfer of wealth is from daughters to their children, that is, inter-generational outcomes post the HSAA reform. The treated women had a greater number of children (Bose & Das, 2020; Tandel et al., 2023) however, the quality of children in terms of for height-for-age z-score indicator[4] worsened and decrease in educational level for children of treated women (Bose & Das, 2017). Therefore, the next generation impact as a result of the inheritance law seems to be unfavorable for the children of the treated women in the reform states.

---

[2] Roy (2015) tests for land gifts to sons which increased post the reform. Since, HSA is valid only for interstate, it could be the case that fathers were taking away their share from the joint property which is inherited under HSA.
[3] Deininger et al., (2013) finds positive impact of likelihood of inheriting land and also increase in educational outcomes for all women. Similar studies (see for example, Bose & Das, 2017; Suteau, 2020) also find evidence for an increase in education years for women exposed to the reform.

Our study is an extension of the extant studies exploring impact of inheritance law and indicators of women empowerment. While most of the studies analysing women autonomy indicators rely on National Family Health Survey (NFHS), we employ India Human Development Survey (IHDS-I) 2004-05 which has a separate module for married women (aged 15–49 years[5]). Moreover, we attempt to incorporate stepwise different dimensions of women empowerment (Kabeer, 1999) starting from access to resources measured using household autonomy and marriage choice. Then, considering agency or control over physical mobility and participation in civil groups such as self-help groups. Ultimately, looking at outcomes or achievements using economic autonomy and intimate partner violence indicators. Thereby, adopting a triangulation approach which provides evidence for any impact of the inheritance law on women's empowerment through changes in the bargaining power. Furthermore, we explore the heterogeneity in the sample based on location, head of household, level of patriarchy in a state, and asset ownership, which are known to influence women's bargaining power.

The landmark amendment in the Hindu Succession Amendment Act (HSAA, 2005) provided all women with an equal right to inherit joint property in the absence of will. Before the federal amendment was passed, five states (Kerala, Andhra Pradesh, Tamil Nadu, Maharashtra, and Karnataka) amendments were implemented between 1976 and 1994 which extended equal inheritance rights for daughters. However, the reform was valid only for unmarried women at the time of reform was passed in their respective states. Thus, creating a quasi-natural experiment for scholars to explore. Given the design of amendment, we employed difference-in-difference (DID) methodology by creating a control and treatment group. The control group consisting of women in the non-reform state and married women in the reform state. Meanwhile, treatment group comprises of unmarried women at the time when reform was passed in their respective state. Further, since HSAA is only applicable to Hindus therefore, we interact our treatment variable with Hindu to get a DID estimator to explore the influence on several dimensions of women empowerment. Additionally, we explore some heterogeneities in our sample based on sector (urban and rural), socio-economic status, level of patriarchy in a state, and gender of the head of the household. Lastly, to check the robustness of our results, we incorporate two falsification tests by analysing our model for non-Hindus for whom HSAA was not applicable and testing our results using a random year of reform.

---

[4] The women exposed to the federal amendment of 2005 show a conflicting pattern of results, that is, decrease in the number of children and improvement in quality of children (height-for-age z-score) (Tandel et al., 2023). The paper discusses about the differences in result for state and federal amendment.
[5] Also termed as "eligible women" module.

Using the Indian Human Development Survey (IHDS-I) (2004–05) dataset and employing the DID technique, the pivotal findings of the study are as follows. First, women exposed to the reform exhibit higher level of physical and civil autonomy with greater say in choosing a husband, but lower household autonomy and lower indicators of intimate partner violence. While post the inheritance reform there are indications of better say in marriage choice and physical mobility, there still exist some gap in realising this change in terms of intra-household bargaining power. Second, in terms of economic autonomy or participation, we found no significant as a result of HSAA. Even after bifurcating our results into agricultural and non-agricultural work, the results remain insignificant. These are in line with the past studies which provide mixed evidence for changes in labour supply (Heath & Tan, 2019; Suteau, 2020). The interaction between institutional legal rights and economic participation is multi-faceted and complex specially in developing countries where legal structures alter before there are changes in the societal and customary rights (Anderson, 2022). Lastly, ascertaining the impact based on socio-economic status we found results are not uniform across quartiles; for instance, economic autonomy is valid only for the middle quartile group. Moreover, household autonomy decreases as we move up the asset quartiles. Further there are evidence that as the level of patriarchy increases in a state, some indicators such as, marriage choice, IPV, and civil autonomy decreases. To sum up, our study highlights the heterogeneity of the impact of inheritance law that are contingent on household and state-level factors.

The rest of the paper is structured as follows. The next section is dedicated to a background of inheritance law followed by discussing the identification strategy, empirical methodology, and descriptive patterns in section three. Section four highlight empirical results and falsification tests. The paper closes with the discussion and conclusion.

## 2. Background

The Hindu Succession Law (HSA) (1956) governs inheritance practice for Hindus which consists of Hindu, Buddhist, Sikh, and Jain[6] dying without a will (intestate succession). This law was applicable to all states except Jammu & Kashmir. The law was an attempt to unify different inheritance practices[7] in India prevailing among different states. It provides the equal right to inherit the personal property (self-acquired property)[8] in case of absence of a will (intestate succession)[9]. In other words, the law extended equal inheritance rights for both sons and daughters[10] for the self-acquired property and not the joint property (ancestral property). As inheritance is a concurrent subject that comes under both the federal and state subjects, states are allowed to make amendments in the inheritance law. Five states; Kerala in 1976, Andhra Pradesh in 1986, Tamil Nadu in 1989, Maharashtra and Karnataka in 1994 passed the state amendments which provided daughters with the equal right to inherit the ancestral property thereby, moving towards an egalitarian inheritance system. Unlike the HSA passed in 1956 which extended inheritance rights for self-acquired property, state amendments extended these rights for the joint property too. However, these inheritance rights were limited to unmarried women at the time of reform in respective states. Apart from five states, other states continue to have male members as the part of coparcenary. However, in 2005, the Hindu Succession Amendment Act (HSAA) was federally implemented all over India (except Jammu & Kashmir) for all Hindu women irrespective of their marital status.

---

[6] The paper uses Hindus throughout the paper for brevity.
[7] Before 1956, inheritance was majorly governed through Dayabhaga for West Bengal and Assam, and Mitakshara for other states. While Dayabhaga system did not made any distinction between joint and self-acquired property,

The state amendment was passed in five states and applicable only to unmarried women in the reform states before it was federally implemented in 2005. Thus, it provided researchers with a quasi-natural experiment to analyze the impact of inheritance law. However, it can be argued that these states are less patriarchal or more female-friendly than other states thus, implemented equal inheritance rights much before it was federally passed. In contrast, Amaral (2017) argues how

---

Mitakshara system had it (Agarwal, 1995). Moreover, the former provided rights for inheriting property to daughters in absence of male heirs and widows while the latter extended no such rights.

[8] The HSA categorises the properties into self-acquired and joint property. The former refers to the acquired property of an individual in a lifetime or inherited from other members of the family excluding father whereas the latter refers to property inherited from the ancestral family upto to three generations.

[9] Most property settlements are governed through HSA as the proportion of people dying without making a will stands at 65 percent for India (Deininger et al., 2013).

[10] The right for inheritance of self-acquired property was also devolved for widows and mothers.

timing of other female-friendly laws such as gender quotas does not coincide with inheritance law. For instance, while inheritance law was implemented in Karnataka in 1994 gender quotas were passed in 1987 followed by Kerala in 1991. Meanwhile, Punjab introduced gender quotas in 1991 and did not pass the amendment in the inheritance law. Moreover, the patriarchy index measured by Singh et al., (2021) comprising of male domination, generational domination, patrilocality, and son preference indicates substantial variation among the five states. This is further strengthened by Anderson (2024) which shows substantial regional differences across India in several women-centric indicators ranging from autonomy, mobility, education, and work. Overall, this indicates that amendment in HSA can be considered as an exogenous shock or a natural experiment to examine the influence of equal inheritance rights on women empowerment as a casual impact.

## 3. Data and Methodology

This section is divided into four sub-sections. The first discusses the data source employed which is followed by explaining the identification strategy and empirical methodology. The last sub-section explores some of the descriptive statistics.

### 3.1 Data Sources

The prime objective of the study is to unveil the impact of Hindu Succession (Amendment) Act (HSAA) on women's empowerment. To do so, the study uses the first round of India Human Development Survey (IHDS-I) (Desai et al., 2005) conducted in 2004-05[11] which is a nationally representative dataset for India. The survey consists of 41,554 households and 215,754 individuals from all states except Andaman and Nicobar and the Lakshadweep thereby, covering 382 districts as per the 2001 national census from a total of 612 districts in India. The dataset covers several dimensions of human development such as education, health, gender relations, fertility, social capital, marriage relations, child-related outcomes, and so on. The questions related to gender relations, health, marital and fertility history were answered by ever-married women (which the survey termed as 'eligible women' aged 15-49 years) whereas questions related to income and social capital were answered by the household heads. The survey also asks questions related to women empowerment such as physical mobility, "Do you have to ask permission of your husband or a senior family member to go to the local health center?". It also includes questions related to household decision-making, participation in the labour force, community perception about intimate partner violence (IPV), and others that could be considered as indicators for measuring women empowerment. Moreover, the time of the survey provides a suitable time frame for analyzing the impact of state amendments of the inheritance law before the HSAA was implemented nationwide in 2005.

---

[11] There are two rounds of IHDS survey, 2004-05 and 2011-12. However, since the first-round year coincides with the federal amendment therefore, in line with the past studies, we also employ the first round.

### 3.2 Identification Strategy

The paper intends to identify the casual impact of inheritance law on women's empowerment in India using the difference-in-difference (DID) technique. Since inheritance law was implemented in some states before it was federally implemented, we can categorize states into reform and non-reform states. Moreover, within the reform states, the state amendments were only applicable to the unmarried women at the time of the reform year. As a result, the control cohort comprises two groups. First, women residing in the non-reform states which are not impacted by the state amendments of the law. Second, women residing in the reform state but are married at the time of the reform year. Meanwhile, the treatment variable in this case is the exposure to the state reform. Therefore, the treatment group comprises women residing in the reform state and unmarried at the time of reform in their respective states. For instance, unmarried women in Tamil Nadu (reform state) in 1989 (the reform year) formed the treatment group whereas married women were part of the control group. It is important to note that law was only valid for households belonging to the religion Hindu, Buddhist, Sikh, and Jain. We consider 16 states for our analysis[12]. A similar DID methodology has been employed by past studies (see for example, Bose & Das, 2017; Rosenblum, 2014; Tandel et al., 2023, among others) to ascertain the inheritance law impact.

Next coming to the dependent variables, women's empowerment, we follow Kabeer's (1999) seminal work on defining women's empowerment by identifying three related dimensions namely, resources (pre-conditions), agency (process), and achievements (outcomes). First, access to resources is defined as control over decision-making or control over "choice" and does not necessarily mean de facto access over physical or monetary assets. For our study, it is measured using the intra-household decision making and choice over choosing the spouse, thereby, measuring household autonomy and marriage choice. Household autonomy is an index containing household decision-making for cooking, purchasing expensive items, and getting cash-in-hand to spend. We consider household autonomy as 1 if respondent has say in at least one decision parameter. Past studies (see for example, Sathar & Kazi, 1997; Kishor, 2000) analysing access to resources uses the dimension of "access" and "control" over the household resources measured using the similar variables relating to whether woman had a say in household expenses, purchases, and other such intra-household financial dynamics. The choice over choosing the partner is measures using "Who chose your husband?", the respondent is given option; "respondent herself, respondent and parents/other relative together, parents or other relatives alone, and other". We test whether treated women have a say in choosing in marriage by considering second option while we test of marriage choice, alone, by considering only first option (the results presented in annexure table A.2).

---

[12] We exclude Kerala which is a reform state as its amendment different than other state amendment. Northeastern states, Jammu Kashmir, and union territories are administered differently hence, are excluded. West Bengal and Assan follows Dayabhaga system are also excluded. The 16 states included in the analysis are as follows; Punjab, Uttaranchal, Haryana, Himachal Pradesh, Rajasthan, Uttar Pradesh, Bihar, Jharkhand, Orissa, Chhattisgarh, Madhya Pradesh, Gujarat, Maharashtra, Andhra Pradesh, Karnataka, and Tamil Nadu.

Second, agency or control explicitly focuses on physical mobility of the women and their presence in the public sphere. For our study, we measure agency using freedom over physical mobility and civil participation which exhibits control over the self-reliance of the respondent. The physical autonomy considers whether women are allowed to venture out to social places (health center, family/friend house, kirana store) without taking permission from husbands or other household members and whether there exists ghunghat (or purdah/pallu/veil) practice[13]. The physical autonomy is 1 if respondent is allowed to venture out without permission for at least one indicator. The civil autonomy variable considers whether women are part of any self-help group (Chatterjee & Desai, 2021). While the household autonomy measures the within household decision making (private sphere), the physical autonomy looks at the decision making outside the home boundaries thereby, measuring control over mobility for women (public sphere).

Lastly, outcomes or achievements is the last leg of women empowerment which consists of both positive and negative outcomes. Within this dimension, we attempt to measure direct evidence of empowerment using paid employment and indicator of intimate partner violence as our indicators. The first indicator measures economic autonomy for women which is measured using women paid work[14] and further bifurcated into agricultural and non-agricultural work indicators. The IPV is a reflection over conflict within the household which is an important indicator to indicate backlash effect of a policy change. Past study analysing institutional changes shows substantial evidence for backlashing through increase in domestic violence (Amaral, 2017; Calvi & Keshar, 2023). The IHDS survey does not measure IPV directly but rather ask the respondent whether "In your community is it usual for husbands to beat their wives?" for different reasons such as "goes out without telling, "neglects house or children", "doesn't cook food properly", and so on. The IPV indicator is 1 if women response is "yes" for at least one question.

---

[13] In India, which is largely a patriarchal society with a 'purdah-pratha' (veil tradition) in several states, mobility provides an opportunity for free movement outside home. Moreover, household mobility parameters like the veil system are linked with economic independence for women (Kandiyoti, 1988).
[14] Women who worked for at least 240 hours in the preceding year were considered employed in paid work (Chatterjee & Desai, 2021).

Overall, by employing three different dimensions of women empowerment and considering a host of indicators, we attempt to access whether influence of inheritance law was uniform across the dimensions. The variables are listed in annexure table A.1

The next sub-section details the empirical strategy used to ascertain the influence of changes in property rights in altering women empowerment.

### 3.3 Empirical Strategy

Using the difference-in-difference strategy followed by the past studies (Bose & Das, 2020; Tandel et al., 2023, among others), the study measures the impact of state amendments passed in Andhra Pradesh, Karnataka, Maharashtra, and Tamil Nadu on women' autonomy[15]. To do so, we form the treatment group as unmarried women in the reform state at the corresponding year of the state amendment. Meanwhile, the control group consists of women in the non-reform states and married women in the reform states at the time of reform year. We estimate the following equation using logit regression[16].

---

[15] The Kerala state amendment was different than the other state amendment hence is excluded from the analysis.

$$Y_{ist} = \beta_0 + \beta_1 \, HSAA_{ist} + \beta_2 \, Hindu_{ist} + \beta_3 \, HSAA_{ist} \times Hindu_{ist} + \beta_4 X_n + \in_{ist} \dots (1)$$

where $i$, $s$, and $t$ represents individual, state, and year respectively. $Y_{ist}$ represents the set of variables measuring women's empowerment described in last sub-section. $HSAA_{ist}$ is a dummy variable equal to 1 if women $i$ is unmarried at the time of reform year and resides in reform state $s$ and 0 otherwise. Since state amendments were only valid for Hindu religion (treated religion), we interact $HSAA_{ist}$ and $Hindu_{ist}$ variables to analyze our difference-in-difference estimator denoted by $\beta_3$. Therefore, our coefficient for interest is $\beta_3$ which depicts the impact of inheritance law on women's empowerment indicators. $X_n$ is a vector of control variables such as education (no school, primary and below, upper primary and secondary, higher secondary, and some college), household size, log of monthly per capita expenditure (MPCE), marriage age, sector (urban or rural), women's age, reform state, caste (brahmin, scheduled caste (SC), schedule tribe (ST), other backward classes (OBC), and others), year of marriage dummies, and state dummies. Though this identification is employed by several past scholars, the approach identifies potentially affected women from the state amendment rather than actual beneficiaries of the state legislation which cannot ascertain due to data unavailability.

Further, we explore our results based on several heterogeneities in the Indian context. First, as social and cultural norms differ among urban and rural areas, inheritance law could have a differential impact on women's autonomy based on region. Therefore, we analyse our results based on sector. Secondly, women empowerment could vary with respect to socio-economic status of the household (Srinivas, 1952) therefore, we bifurcate our sample into different asset quartiles as an indicator for economic status (Tandel et al., 2023). Third, extant literature discusses about the north-south dichotomy in terms of differences in indicators of women autonomy (see for example, Dyson & Moore, 1983; Anderson, 2024). Therefore, we incorporate the level of patriarchy at the state level to understand how impact of institutional amendment interacts with state-level dynamics in altering women-related indicators. To do so, we use the patriarchy index[17] developed by Singh et al. (2021) at a state-level and run a triple-difference estimation by interacting our difference-in-difference estimator with patriarchy index. Lastly, based on the gender of head of the household, we analyse whether female headed households could reap some additional benefits in improving women empowerment than male-headed households. We analysis this impact using the triple-difference estimation.

---

[16] We also estimate using the ordered logistic regression for physical and household autonomy and result exhibit similar patterns.

To further test the robustness of our results we include two falsification tests. First, considering influence of inheritance rights on non-Hindu which were not part of the inheritance law. Second, checking our results by considering random year of reform before any of the state amendment were passed.

### 3.4 Descriptive Statistics

After formulating the empirical methodology in the last sub-section, we discuss some summary statistics for the listed variables in this sub-section. In table 1, we present the descriptive statistics for the independent variables for all states (column 1), non-reform states (column 2), reform states (column 3), and within the reform states, treated (column 4) and control group (column 5). Overall, there exist significant differences between the characteristics of treated women and non-treated women. In the reform states, the mean age of marriage and education level is higher than the non-reform state. However, the pattern seems to be driven by the treated women in the reform state where the average age of marriage is 18 years whereas it is approximately 16 years for the non-treated women (16.73 years) and for the non-reform states (16.94 years). Similarly, while women in the treated group with at least primary education is 64.92 percent, in the control group and non-reform state, it is 46.89 percent and 43.61 percent respectively. Thus, it seems that treated women tend to be more educated on an average than non-treated women. Past studies (see for example, Deininger et al., 2019; Roy, 2015) also indicates that as a result of HSAA, treated women have experienced an increase human capital through additional educational years and expenses.

---

[17] To scale the value, we divide patriarchy index provided by Singh et al. (2021) with its range (maximum-minimum).

**<Table 1 here>**

Next, for our dependent variables we present the descriptive statistics in *table 2*. Almost all the autonomy indicators (civil, physical, household, and economic) are substantially higher for the women in the reform state as compared to the non-reform state. Moreover, indicator for intimate partner violence is lower for the women in the treated cohort (0.915) as compared to the control group (0.925). Even for the indicator of who chose the husband, treated women (0.63) in the reform states have a greater autonomy in choosing husband alone than in non-reform states (0.36). Since descriptive statistics present the averages thus, we control for several individual-level and household-level factors in the regression analysis which are presented in the subsequent section.

**<Table 2 here>**

## 4. Findings

The section first discusses the main regression findings followed by exploring the falsification checks.

### 4.1 Regression Results

We present the regression results for the first dimension of women empowerment; access to resources, estimated using equation (1) in *table 3*. The access to resources is measured using index for household autonomy and say in choosing one's husband (marriage choice). Column (2) indicates, women exposed to state reform are 47 percent more likely to have a say in choosing her husband than women in the control group. However, when we test whether she can choose her husband alone, the coefficient indicates a significantly negative relationship (annexure table A.2). Furthermore, past studies (see for example, Rexer, 2022) indicate a positive relationship between expected income realization and choice of husband in terms of wealth. In contrast, our results[18] indicates that treated women are marry down in terms of economic status. Thus, the analysis highlights that, post the reform, women have gained bargaining power in having some say choosing their partner which could have further translate to marrying down economically. Moreover, the pattern persists across urban and rural sectors. The second indicator, the household autonomy index which comprises of indicators related to intra-household decision making; indicates that treated women are likely to have low household autonomy as a result of inheritance rights. While the first indicator of marriage choice pertains to decision-making in their natal family while household autonomy of women is an indicator for bargaining power in the husband's family. Therefore, while marriage choice has substantially increased post the reform, intra-household decision making which interacts with relationship with spouse and in-laws are unchanged.

---

[18] IHDS provides us with information regarding whether the economic status of a natal family is better/same/worse off than the husband's family. The results are provided in the annexure table A.2 where we also incorporate marriage in choice (say in choosing husband) as a control variable.

**<Table 3 here>**

Coming to the next dimension of women empowerment, agency or control over movement which is measured using physical autonomy index and civil indicator (part of self-help group (SHG)) (table 4). The result indicates post the reform, women in the treatment group are significantly more likely to gain physical autonomy irrespective of living in urban or rural area and are a part of a SHG. The results are in line with the past studies (see for example, Mookerjee, 2019, Roy, 2008) that indicates positive impact of HSAA on freedom in mobility for the treated cohort.

**<Table 4 here>**

Lastly, we focus on outcomes or achievements indicators using economic participation and intimate partner violence (IPV) (table 5). The results indicate a significant decrease in IPV as a result of the reform thereby, providing evidence against backlashing effect in terms of domestic violence. However, the pattern is observed only for the urban sector. In terms of economic autonomy, there exists a positive but insignificant increase in paid work for women in the treated group. Further exploring if there exist any differences when we bifurcate paid work into agricultural and non-agricultural work[19], we still do not find any evidence for increase in labour supply. Past studies exhibit mixed evidence for changes in labor supply. For instance, on one hand Suteau's (2020) analysis indicate no significant increase in labour supply for women exposed to the reform whereas Heath & Tan (2019) show increase in professional jobs. The intersection of institutional changes and economic change is webbed into complex structures which are further hindered by restrictive social norms. Therefore, the channel through which institutional change such as inheritance rights could impact economic autonomy can be twofold. First, post the inheritance rights, women outside options increases thereby, leading to rise in non-labor income which could decrease willingness to participate in the labor force. Secondly, according to bargaining model, a rise in women outside option increases her bargaining power thus, could provide her with greater control to participate in the labor market. Therefore, a favorable institutional amendment towards women can both increase and decrease the economic participation for women and is contingent upon several unobservable factors.

---

[19] The estimation results for agricultural and non-agricultural work are available in the annexure in table A.3.

**\<Table 5 here\>**

In the above results, we look at the impact on an average household. However, the impact of the reform could vary based on household factors such as, the socio-economic status of the family or gender of the head of the household; and the state-level factors such as restrictive social norms indicated through level of patriarchy. We attempt to further analyze our results based on these heterogeneities.

While we cannot directly measure the economic status of the household, we can use asset ownership as a proxy. Therefore, we create four quartiles of asset ownership to access how the impact of HSAA varies with respect to socio-economic dynamics[20]. The impact of reform varies considerably based on asset quartiles (table 6). Overall, we found three pivotal findings. First, women in the treated cohort have gained substantial say in choosing their husband, especially in households belonging in first and second quartiles. However, the household autonomy is significantly low at the fourth quartile which could indicate that, women in households who have better socio-economic status are more likely to experience decrease in intra-household bargaining power. Secondly, in the case of civil autonomy, while the third and fourth quartile experiences positive and significant impact of inheritance law, lower quartile households experience a negative likelihood of being a part of SHGs. Lastly, for economic autonomy, only the third quartile group experiences a positive and significant impact for the treatment. We show in the annexure *(table A.4)* that the economic autonomy rises in the non-agricultural work for the treated women whereas it is insignificant for agricultural work. In contrast, likelihood of IPV decreases for women in households with greater socio-economic status. Hence, overall highlighting that the impact of state amendments is not uniform across the wealth distribution of households based on the asset ownership quartiles.

---

[20] Tandel et al., (2023) uses a similar strategy to analyze the quality-quantity trade-off of children as a result of HSAA for various asset ownership quartiles.

**<Table 6 here>**

Given the substantial variation among Indian states based on cultural and social norms, it would be interesting to explore how impact of institutional amendments varies among states based on level of patriarchy. Therefore, we incorporate patriarchy index at the state level. The patriarchy index[21] developed by Singh et al., (2021) uses four dimensions; male domination, generational domination, patrilocality, and preference for son. Using the triple-difference method, we interact, treated women, Hindu, and patriarchy index. The results presented in table 7, indicates that states with higher levels of patriarchy exhibit lower or even negative impact of inheritance law on treated women. For instance, while overall we observed women exposed to reform experienced greater say in the choosing their husband however, when we interact with patriarchy index, the coefficient for triple interaction exhibits negative and significant relationship. Even for IPV, there exist positive coefficient though insignificant unlike the pattern exhibited in the above results which showed a decrease IPV as a result of state amendments of inheritance law. Thereby, highlighting a uniform institutional change render differential (or even opposite) results when interacted with state-level forces.

---

[21] For our study, we use patriarchy index (Singh et al., 2021) developed using National Family Health Survey (NFHS-3) (2005-06) which fits closest to IHDS (2004-05) year.

**<Table 7 here>**

Lastly, we check for the influence of gender of head of household on women empowerment indicator as a result of HSAA. It is expected that female-headed households exhibit greater autonomy and bargaining power, hence, should have greater positive impact post the reform. By employing triple-difference estimation, the results in table 8 highlights no substantial difference in household with female as the head. The coefficient for female headed household variable show better outcomes for almost all the indicators thus, the impact of reform might not have bought any considerable change in their empowerment.

**<Table 8 here>**

In a nutshell, our analysis highlights how institutional amendments impact different dimensions of women empowerment. As highlighted by Anderson (2022; 2024) different indicators of women empowerment might not move in tandem with each other which is highlighted in this case, where there exists evidence for improvement in physical and civil autonomy that is, part of public sphere but not within the household which is part of private sphere. While not all dimensions move in concert however, as a result of HSAA, indicators such as physical, civil, marriage choice, and IPV exhibit favorable changes.

## 4.2 Falsification Tests

We incorporate two falsification tests. Ideally, without the reform, the indicators should exhibit similar trends between the treated and control cohort. First, we estimate our results for non-Hindus for whom HSAA was not valid hence, we should not find any significant impact on women empowerment. We restrict the sample to the non-Hindus and estimate equation analogous to equation (1). Second, we test our analysis using a random year of reform, that is, 1975 which is one year prior to the Kerala's amendment year. The treatment variable (HSAA) considers the value 1 if women $i$ resides in reform state and is unmarried at the time of false reform year (1975).

**<Table 9 here>**

The results for the first falsification tests are presented in table 9 in Panel A. We restrict the sample to non-Hindus. For the dimensions of women empowerment, the results indicate either an insignificant relationship or a negative impact on non-Hindus. For instance, marriage choice, physical, civil, and IPV indicates opposite impact on non-Hindus. The result could also be driven by the religion group included in the category of non-Hindus which might have low women empowerment levels. The results provide evidence that impact of reform was particular to Hindu women for whom the reform was directed.

The second falsification tests consider a random reform year that is, 1975 when no state amendment was passed. Apart from the physical autonomy indicators, other indicators of women empowerment possess an insignificant impact on falsely treated women. Overall, other indicators such as, IPV, marriage choice, civil, and economic autonomy variables are not statistically significant. Therefore, lending support to our analysis that inheritance reform has a positive impact on women empowerment.

## 5. Conclusions

Our study analyses the impact of the Hindu Succession Amendment Act (HSAA) on women's empowerment indicators. To do so, we employ the India Human Development Survey (IHDS-1) 2004–05 and adopt Kabeer's (1999) triangulation approach of accessing women empowerment, measuring access to resources (household and marriage choice), agency (physical and civil autonomy), and outcomes (economic autonomy and intimate partner violence). Since HSAA was amended in five states between 1976 and 1994 before it was implemented nation-wide in India, we form our control and treatment groups and employ the difference-in-difference technique. Our study focusing on several dimensions of women's empowerment provides positive evidence for an increase in, say in choosing their husband (marriage choice), physical mobility, and participation in self-help groups (civil autonomy) with improvements in IPV indicators. In contrast, household autonomy measured using indicators related to say in intra-household decision-making exhibits a significantly negative relationship. Thus, overall highlighting that different dimensions of women empowerment does not necessarily move in alliance with each other.

The mechanism through which institutional change impact women are webbed into complex dynamics both at a household level and societal level which might not be directly linked to each other. Moreover, they interact with several heterogeneities present within the household and society such as, economic status of the household, state level dynamics of social norms, and so on. Thus, we further incorporate economic status based on asset ownership quartiles and patriarchy index at the state level to explore some of the heterogeneities. The results indicates that the household autonomy weakens as we move up the asset ownership quartiles, in other words, as households get richer, household autonomy decreases further for women exposed to the reform. Additionally, being a part of SHG or decrease in IPV indicators are significant for households in third and fourth quartile. Thus, overall highlighting impact of inheritance rights on women empowerment is not uniform across different economic status of the households. Next, while interacting our results with level of patriarchy, we found, women in states with higher patriarchy level tend to have less impact on women empowerment. In a nutshell, our results indicate that changing inheritance rights in favour of women leads to positive changes in indicators of women empowerment to some extent however it is contingent on economic status, sector (urban or rural), and state-level patriarchal norms. Moreover, further research is required in analysing how despite several institutional amendments in terms of stretching political rights or economic rights, is not resulting in altering economic participation for women.

In terms of policy implications, the study provides how an institutional change by extending economic rights via equal inheritance impact different dimensions of women empowerment. One key aspect to remember in this context is that the impact is not homogenous and is mediated by various other factors affecting the within household dynamics. First, we can say that right based approach to provide gender equality can have long lasting effects but will be more effective when done in tandem with each other instead of being done in isolation. As is observed in our results though the autonomy of women increases in states with equal inheritance rights, this effect is less visible in higher income households. Thus along with HSAA, may be a series of reforms for women empowerment through affirmative action in education, labour market and political participation in combination will have long lasting and more pronounced effect on outcomes for women. As we find in our studies, though there has been a substantial average improvement in outcomes for women on account of HSAA. This effect is not same for all and remains miles away from full empowerment.

| | All States | Non-Reform States | Reform States | | |
|---|---|---|---|---|---|
| | | | **All** | **Treated** | **Control** |
| | **(1)** | **(2)** | **(3)** | **(4)** | **(5)** |
| **Caste** | | | | | |
| **Brahmin** | 0.050 | 0.065 | 0.021 | 0.020 | 0.021 |
| **OBC** | 0.439 | 0.414 | 0.488 | 0.487 | 0.488 |
| **SC** | 0.218 | 0.219 | 0.215 | 0.214 | 0.216 |
| **ST** | 0.079 | 0.088 | 0.062 | 0.050 | 0.071 |
| **Other** | 0.214 | 0.213 | 0.214 | 0.229 | 0.204 |
| | | | | | |
| **Hindu** | 0.113 | 0.113 | 0.114 | 0.113 | 0.115 |
| **Non-Hindu** | 0.887 | 0.887 | 0.886 | 0.887 | 0.885 |
| | | | | | |
| **Women's education** | | | | | |
| **Illiterate** | 0.527 | 0.564 | 0.455 | 0.351 | 0.531 |
| **Primary and below** | 0.161 | 0.161 | 0.161 | 0.144 | 0.174 |
| **Upper Primary and Secondary** | 0.234 | 0.199 | 0.302 | 0.372 | 0.251 |
| **Higher Secondary** | 0.034 | 0.034 | 0.036 | 0.059 | 0.019 |
| **Some College** | 0.044 | 0.043 | 0.045 | 0.074 | 0.025 |
| | | | | | |
| **Sector** | | | | | |
| **Rural** | 0.678 | 0.708 | 0.619 | 0.582 | 0.646 |
| **Urban** | 0.323 | 0.292 | 0.381 | 0.418 | 0.354 |
| | | | | | |
| **Age** | 32.994 | 33.082 | 32.822 | 26.230 | 37.636 |
| **Marriage Age** | 17.117 | 16.941 | 17.460 | 18.451 | 16.737 |
| **Household Size** | 6.592 | 6.964 | 5.868 | 5.975 | 5.791 |
| **MPCE (log)** | 6.421 | 6.373 | 6.513 | 6.522 | 6.506 |
| | | | | | |
| **Observations** | 56,431 | 37,262 | 19,169 | 8,091 | 11,078 |

**Table 1: Descriptive Statistics for Independent Variables**

Note: Probability weights are used.
Source: Authors' compilation using IHDS - I (2004-05)

**Table 2: Descriptive Statistics for Dependent Variables**

| | All States | Non-Reform States | Reform States | | |
|---|---|---|---|---|---|
| | | | All | Treated | Control |
| | (1) | (2) | (3) | (4) | (5) |
| **Household Autonomy** | 0.982 | 0.979 | 0.989 | 0.981 | 0.995 |
| *Say in Cooking Decision* | 0.948 | 0.938 | 0.966 | 0.948 | 0.979 |
| *Say in purchase of expensive item* | 0.744 | 0.726 | 0.779 | 0.726 | 0.818 |
| *Cash in hand to spend* | 0.868 | 0.865 | 0.873 | 0.857 | 0.884 |
| | | | | | |
| **Choice to choose husband** | | | | | |
| *Herself* | 0.041 | 0.036 | 0.05 | 0.063 | 0.041 |
| *Respondent and Parents together* | 0.403 | 0.368 | 0.471 | 0.431 | 0.5 |
| *Parents alone* | 0.553 | 0.593 | 0.476 | 0.502 | 0.456 |
| *Other* | 0.003 | 0.003 | 0.003 | 0.004 | 0.003 |
| | | | | | |
| **Physical Autonomy** | 0.666 | 0.533 | 0.924 | 0.926 | 0.922 |
| *Permission for Health Centre* | 0.24 | 0.21 | 0.298 | 0.27 | 0.319 |
| *Permission for Friends/relative house* | 0.24 | 0.213 | 0.294 | 0.286 | 0.299 |
| *Permission for Kirana Store* | 0.398 | 0.358 | 0.478 | 0.482 | 0.475 |
| *Purdah Practice* | 0.46 | 0.277 | 0.815 | 0.83 | 0.804 |
| | | | | | |
| **Civic Autonomy** | 0.101 | 0.056 | 0.189 | 0.191 | 0.188 |
| | | | | | |
| **Economic Autonomy** | 0.326 | 0.323 | 0.333 | 0.288 | 0.365 |
| Agricultural Work | 0.092 | 0.067 | 0.14 | 0.115 | 0.158 |
| Non-Agricultural Work | 0.026 | 0.022 | 0.034 | 0.028 | 0.038 |
| | | | | | |
| **Intimate Partner Violence** | 0.917 | 0.915 | 0.921 | 0.915 | 0.925 |
| *Goes out without telling* | 0.412 | 0.399 | 0.438 | 0.39 | 0.474 |
| *Her natal family does not give expected money, jewellery, or other items* | 0.289 | 0.247 | 0.369 | 0.345 | 0.386 |
| *She neglects house or children* | 0.363 | 0.318 | 0.451 | 0.427 | 0.468 |
| *She doesn't cook food properly* | 0.308 | 0.28 | 0.362 | 0.332 | 0.383 |
| *He suspects her of having relations with other men* | 0.903 | 0.903 | 0.903 | 0.892 | 0.91 |
| **Observations** | 56,431 | 37,262 | 19,169 | 8,091 | 11,078 |

Note: Probability weights are used.

Source: Authors' compilation using IHDS - I (2004-05)

**Table 3: Regression estimates of impact of HSAA on Household autonomy and Marriage Choice**

| Dependent Variable | (1)<br>Household Autonomy | (2)<br>Marriage Choice |
|---|---|---|
| **HSAA* Hindu** | -0.870*** | 0.388*** |
| | (0.335) | (0.0821) |
| **HSAA** | 0.119 | -0.534*** |
| | (0.354) | (0.0825) |
| **Hindu** | 0.252** | -0.101*** |
| | (0.106) | (0.0353) |
| **Observations** | 55,452 | 56,402 |
| **Only Urban** | | |
| **HSAA* Hindu** | 0.266 | 0.221** |
| | (0.400) | (0.109) |
| **HSAA** | -0.611 | -0.517*** |
| | (0.430) | (0.110) |
| **Hindu** | -0.100 | -0.0673 |
| | (0.169) | (0.0518) |
| **Observations** | 14,297 | 18,185 |
| **Only Rural** | | |
| **HSAA* Hindu** | - | 0.562*** |
| | | (0.133) |
| **HSAA** | -0.881*** | -0.625*** |
| | (0.228) | (0.133) |
| **Hindu** | 0.336** | -0.120** |
| | (0.143) | (0.0501) |
| **Observations** | 34,919 | 38,207 |

Note: All regression includes reform state, education, age, marriage age, household size, log of MPCE, urban dummy, caste, year of marriage dummies, and state dummies as control variables.
Standard errors in parentheses.
*** p<0.01, ** p<0.05, * p<0.1
Source: Authors' calculation using IHDS - I (2004-05)

**Table 4: Regression estimates of impact of HSAA on Physical and Civil autonomy**

| Dependent Variable | (1)<br>Physical Autonomy | (2)<br>Civil Autonomy |
|---|---|---|
| **HSAA* Hindu** | 2.219*** | 0.417*** |
|  | (0.102) | (0.130) |
| **HSAA** | -1.594*** | -0.368*** |
|  | (0.0918) | (0.132) |
| **Hindu** | 0.533*** | 0.0835 |
|  | (0.0361) | (0.0699) |
| **Observations** | 56,419 | 56,361 |
| **Only Urban** |  |  |
| **HSAA* Hindu** | 1.874*** | 0.731*** |
|  | (0.148) | (0.272) |
| **HSAA** | -1.378*** | -1.023*** |
|  | (0.133) | (0.278) |
| **Hindu** | 0.554*** | 0.518*** |
|  | (0.0570) | (0.131) |
| **Observations** | 18,185 | 18,076 |
| **Only Rural** |  |  |
| **HSAA* Hindu** | 2.557*** | 0.110 |
|  | (0.150) | (0.159) |
| **HSAA** | -1.829*** | 0.00190 |
|  | (0.138) | (0.161) |
| **Hindu** | 0.463*** | -0.0576 |
|  | (0.0486) | (0.0845) |
| **Observations** | 38,224 | 38,169 |

Note: All regression includes reform state, education, age, marriage age, household size, log of MPCE, urban dummy, caste, year of marriage dummies, and state dummies as control variables.
Standard errors in parentheses.
*** p<0.01, ** p<0.05, * p<0.1
Source: Authors' calculation using IHDS - I (2004-05)

**Table 5: Regression estimates of impact of HSAA on Economic autonomy and IPV**

| Dependent Variable | (1)<br>Economic Autonomy | (2)<br>IPV |
|---|---|---|
| **HSAA* Hindu** | 0.140 | -0.632*** |
| | (0.104) | (0.160) |
| **HSAA** | -0.219** | 0.697*** |
| | (0.104) | (0.160) |
| **Hindu** | 0.361*** | -0.0919 |
| | (0.0379) | (0.0607) |
| **Observations** | 56,431 | 56,360 |
| **Only Urban** | | |
| **HSAA* Hindu** | 0.220 | -0.703*** |
| | (0.165) | (0.182) |
| **HSAA** | -0.256 | 0.738*** |
| | (0.166) | (0.185) |
| **Hindu** | 0.257*** | 0.189** |
| | (0.0693) | (0.0759) |
| **Observations** | 18,196 | 18,170 |
| **Only Rural** | | |
| **HSAA* Hindu** | -0.200 | 0.552 |
| | (0.135) | (0.371) |
| **HSAA** | 0.444*** | -0.210* |
| | (0.0459) | (0.113) |
| **Hindu** | 0.0990 | -0.592 |
| | (0.135) | (0.372) |
| **Observations** | 38,232 | 37,802 |

Note: All regression includes reform state, education, age, marriage age, household size, log of MPCE, urban dummy, caste, year of marriage dummies, and state dummies as control variables.
Standard errors in parentheses.
*** $p<0.01$, ** $p<0.05$, * $p<0.1$
Source: Authors' calculation using IHDS - I (2004-05)

**Table 6: Regression estimates of impact of HSAA on women's empowerment indicators (Asset Ownership Quartile-wise)**

| Dependent Variable | (1) First | (2) Second | (3) Third | (4) Fourth |
|---|---|---|---|---|
| *(i) Household* | | | | |
| **HSAA* Hindu** | 0.418 | 0.458 | - | -1.300* |
| | (0.790) | (0.566) | - | (0.675) |
| **Observations** | 14,766 | 6,195 | 12,101 | 8,177 |
| *(ii) Marriage Choice* | | | | |
| **HSAA* Hindu** | 0.884*** | 1.018*** | 0.226* | 0.0846 |
| | (0.234) | (0.242) | (0.128) | (0.156) |
| **Observations** | 15,918 | 9,868 | 16,758 | 13,809 |
| *(iii) Physical* | | | | |
| **HSAA* Hindu** | 2.014*** | 2.928*** | 2.003*** | 2.221*** |
| | (0.276) | (0.282) | (0.158) | (0.208) |
| **Observations** | 15,930 | 9,895 | 16,764 | 13,793 |
| *(iv) Civil* | | | | |
| **HSAA* Hindu** | -0.483* | 0.435 | 0.674*** | 0.926** |
| | (0.283) | (0.297) | (0.210) | (0.365) |
| **Observations** | 15,210 | 9,732 | 16,712 | 13,642 |
| *(v) Economic* | | | | |
| **HSAA* Hindu** | -0.0653 | 0.0738 | 0.349** | -0.0274 |
| | (0.227) | (0.225) | (0.170) | (0.244) |
| **Observations** | 15,935 | 9,914 | 16,768 | 13,805 |
| *(vi) IPV* | | | | |
| **HSAA*Hindu** | 0.488 | - | -0.687*** | -0.663** |
| | (0.414) | | (0.248) | (0.271) |
| **Observations** | 15,504 | 9,008 | 16,664 | 13,703 |

Note: All regression includes reform state, education, age, marriage age, household size, log of MPCE, urban dummy, caste, year of marriage dummies, and state dummies as control variables.

Standard errors in parentheses.

*** p<0.01, ** p<0.05, * p<0.1

Source: Authors' calculation using IHDS - I (2004-05)

**Table 7: Regression estimates of impact of HSAA on women empowerment indicators (Patriarchy Index)**

| Dependent Variable | (1) Household | (2) Marriage Choice | (3) Physical | (4) Civil | (5) Economic | (6) IPV |
|---|---|---|---|---|---|---|
| **HSAA* Hindu*Patriarchy Index** | -3.478** | -1.146*** | 2.861*** | 0.507 | 0.647 | 0.157 |
| | (1.682) | (0.341) | (0.544) | (0.514) | (0.462) | (0.556) |
| **HSAA** | -1.228 | -1.949*** | 2.369*** | -0.000557 | 0.163 | -1.228 |
| | (2.319) | (0.421) | (0.508) | (0.618) | (0.576) | (2.319) |
| **Hindu** | -2.348* | -0.603** | 5.458*** | 1.025*** | 0.684** | -2.348* |
| | (1.415) | (0.256) | (0.335) | (0.389) | (0.307) | (1.415) |
| **Patriarchy Index** | -1.765 | 0.597*** | 1.120*** | -0.489* | 0.151 | -1.765 |
| | (1.107) | (0.182) | (0.245) | (0.283) | (0.218) | (1.107) |
| **HSAA* Hindu** | 4.079* | 1.832*** | -2.208*** | -0.312 | -0.702 | 4.079* |
| | (2.308) | (0.437) | (0.718) | (0.637) | (0.592) | (2.308) |
| **Observations** | 55,452 | 56,402 | 56,419 | 56,361 | 56,431 | 56,360 |

Note: All regression includes reform state, education, age, marriage age, household size, log of MPCE, urban dummy, caste, year of marriage dummies, and state dummies as control variables.
Standard errors in parentheses.
*** p<0.01, ** p<0.05, * p<0.1
Source: Authors' calculation using IHDS - I (2004-05)

**Table 8: Regression estimates of impact of HSAA on women empowerment indicators for Female Headed Households**

| Dependent Variable | (1) Household | (2) Marriage Choice | (3) Physical | (4) Civil | (5) Economic | (6) IPV |
|---|---|---|---|---|---|---|
| **HSAA* Hindu*Female Headed Households** | 0.701 | 0.131 | 1.730*** | 0.793* | -0.432 | -1.090 |
| | (0.928) | (0.324) | (0.650) | (0.408) | (0.336) | (0.672) |
| **HSAA** | 0.402 | -0.513*** | -0.235* | -1.519*** | -0.274** | 0.621*** |
| | (0.400) | (0.0852) | (0.136) | (0.0945) | (0.110) | (0.164) |
| **Hindu** | 0.249** | -0.106*** | 0.140* | 0.552*** | 0.355*** | -0.103 |
| | (0.108) | (0.0367) | (0.0747) | (0.0376) | (0.0398) | (0.0640) |
| **Female Headed Households** | 0.703 | -0.279** | 0.624*** | 0.812*** | 0.277** | -0.369** |
| | (0.515) | (0.114) | (0.193) | (0.118) | (0.114) | (0.176) |
| **HSAA* Hindu** | -0.992*** | 0.383*** | 0.292** | 2.156*** | 0.186* | -0.544*** |
| | (0.385) | (0.0852) | (0.135) | (0.105) | (0.110) | (0.165) |
| **Observations** | 55,452 | 56,402 | 56,361 | 56,419 | 56,431 | 56,360 |

Note: All regression includes reform state, education, age, marriage age, household size, log of MPCE, urban dummy, caste, year of marriage dummies, and state dummies as control variables.

Standard errors in parentheses.

*** p<0.01, ** p<0.05, * p<0.1

Source: Authors' calculation using IHDS - I (2004-05)

**Table 9: Falsification Tests: Regression estimates of impact of HSAA on women's empowerment indicators**

| Dependent Variable | (1)<br>Household | (2)<br>Marriage Choice | (3)<br>Physical | (4)<br>Civil | (5)<br>Economic | (6)<br>IPV |
|---|---|---|---|---|---|---|
| **(A) Falsification Test 1 (Only Non-Hindus)** | | | | | | |
| **HSAA** | -14.03 | -0.518*** | -0.310** | -0.468** | 0.0443 | 0.460** |
| | (515.2) | (0.119) | (0.125) | (0.237) | (0.145) | (0.223) |
| **Observations** | 3,489 | 6,343 | 6,397 | 5,804 | 6,397 | 5,844 |
| **(B) Falsification Test 2 (Reform Year 1975)** | | | | | | |
| **HSAA*Hindu** | -1.213*** | - | 2.743*** | -0.0459 | 0.0822 | -0.206* |
| | (0.330) | - | (0.0744) | (0.131) | (0.0735) | (0.115) |
| **HSAA** | 1.188** | 0.527 | -1.635*** | 0.227 | -0.0959 | 0.181 |
| | (0.531) | (0.325) | (0.136) | (0.172) | (0.107) | (0.197) |
| **Hindu** | 0.321*** | - | 0.131*** | 0.241** | 0.351*** | -0.118* |
| | (0.107) | - | (0.0376) | (0.112) | (0.0430) | (0.0696) |
| **Observations** | 55,452 | 6,343 | 56,419 | 56,361 | 56,431 | 56,360 |

Note: All regression includes reform state, education, age, marriage age, household size, log of MPCE, urban dummy, caste, year of marriage dummies, and state dummies as control variables.
Standard errors in parentheses.
*** p<0.01, ** p<0.05, * p<0.1
Source: Authors' calculation using IHDS - I (2004-05)

**Table A.1: Variables used for measuring Women Empowerment**

| Dimension of Women Empowerment | Indicator | Variables |
|---|---|---|
| **Access to Resources** | Household Autonomy Index | *Say in Cooking Decision* |
| | | *Say in purchase of expensive item* |
| | | *Cash in hand to spend* |
| | Marriage Choice | *Who chose your husband?* |
| **Agency** | Physical Autonomy Index | *Permission for Health Centre* |
| | | *Permission for Friends/relative house* |
| | | *Permission for Kirana Store* |
| | | *Purdah Practice* |
| | Civil Autonomy | *Women belong to any self-help group (SHG)* |
| **Outcomes** | Economic Autonomy | *Paid Employment (Agricultural and non-agricultural)* |
| | Intimate Partner Violence | *Goes out without telling* |
| | | *Her Natal Family does not give expected money, jewellery, or other items* |
| | | *Neglects the house or the children* |
| | | *Does not cook food properly* |
| | | *He suspects her of having relations with other men* |

**Table A. 2: Regression estimates of impact of HSAA on Marriage Choice and Natal Family Status**

| Dependent Variable | (1) Marriage Choice (Alone) | (2) Natal Family Status (Same) | (3) Natal Family Status (Better) | (4) Natal Family Status (Worse) |
|---|---|---|---|---|
| HSAA* Hindu | -0.467*** | -0.116 | 0.304*** | -0.267** |
| | (0.150) | (0.0867) | (0.106) | (0.130) |
| HSAA | 0.605*** | 0.00638 | -0.299*** | 0.598*** |
| | (0.153) | (0.0869) | (0.105) | (0.132) |
| Hindu | -0.0470 | 0.0858** | -0.110*** | -0.192*** |
| | (0.0843) | (0.0345) | (0.0386) | (0.0603) |
| Observations | 56,284 | 56,329 | 56,317 | 56,236 |
| **Only Urban** | | | | |
| HSAA* Hindu | 0.0220 | -0.154 | 0.248* | -0.316* |
| | (0.200) | (0.116) | (0.139) | (0.186) |
| HSAA | 0.0751 | -0.0629 | -0.0407 | 0.633*** |
| | (0.207) | (0.117) | (0.138) | (0.191) |
| Hindu | -0.430*** | 0.216*** | -0.260*** | -0.117 |
| | (0.117) | (0.0523) | (0.0577) | (0.0969) |
| Observations | 18,021 | 18,162 | 18,162 | 18,143 |
| **Only Rural** | | | | |
| HSAA* Hindu | -1.168*** | -0.138 | 0.512*** | -0.305 |
| | (0.237) | (0.141) | (0.184) | (0.194) |
| HSAA | 1.378*** | 0.112 | -0.639*** | 0.628*** |
| | (0.240) | (0.141) | (0.183) | (0.196) |
| Hindu | 0.342*** | -0.0224 | -0.0292 | -0.170** |
| | (0.132) | (0.0482) | (0.0551) | (0.0810) |
| Observations | 38,096 | 38,157 | 38,145 | 38,071 |

Note: All regression estimate control for reform state, education, age, marriage age, household size, log of MPCE, urban dummy, caste, year of marriage dummies, and state dummies. Further, column (2) – (4) also includes say in choosing husband as a control variable.
Standard errors in parentheses.
*** p<0.01, ** p<0.05, * p<0.1
Source: Authors' calculation using IHDS - I (2004-05)

**Table A. 3: Regression estimates of impact of HSAA on economic autonomy (agricultural and non-agricultural work)**

| Dependent Variable | (1) Agricultural Work | (2) Non-Agricultural Work |
|---|---|---|
| **HSAA* Hindu** | -0.0415 | 0.391 |
| | (0.175) | (0.238) |
| **HSAA** | -0.0751 | -0.403* |
| | (0.176) | (0.238) |
| **Hindu** | 0.353*** | -0.484*** |
| | (0.0813) | (0.0921) |
| **Observations** | 56,427 | 56,390 |
| **Only Urban** | | |
| **HSAA* Hindu** | -0.767 | -0.343 |
| | (0.512) | (0.353) |
| **HSAA** | -0.767 | -0.343 |
| | (0.512) | (0.353) |
| **Hindu** | 0.536** | -0.158 |
| | (0.236) | (0.145) |
| **Observations** | 14,971 | 18,181 |
| **Only Rural** | | |
| **HSAA* Hindu** | -0.0726 | 0.082 |
| | (0.190) | (0.337) |
| **HSAA** | 0.368*** | -0.689*** |
| | (0.0875) | (0.124) |
| **Hindu** | 0.336** | -0.120** |
| | (0.143) | (0.0501) |
| **Observations** | 38,232 | 38,198 |

Note: All regression estimate control for reform state, education, age, marriage age, household size, log of MPCE, urban dummy, caste, year of marriage dummies, and state dummies. Standard errors in parentheses.
*** p<0.01, ** p<0.05, * p<0.1
Source: Authors' calculation using IHDS - I (2004-05)

**Table A.4: Regression estimates of impact of HSAA on agricultural and non-agricultural work (Asset Ownership quartile-wise)**

| | (1) | (2) | (3) | (4) |
|---|---|---|---|---|
| **Dependent Variable** | **First** | **Second** | **Third** | **Fourth** |
| *Agricultural Work* | | | | |
| **HSAA* Hindu** | -0.174 | -0.0860 | -0.0814 | |
| | (0.269) | (0.341) | (0.345) | |
| **Observations** | 15,928 | 9,780 | 16,711 | 9,863 |
| *Non-agricultural Work* | | | | |
| **HSAA* Hindu** | 0.445 | 0.396 | 0.953** | -0.234 |
| | (0.633) | (0.442) | (0.375) | (0.770) |
| **Observations** | 15,850 | 9,331 | 15,678 | 10,889 |

Note: All regression includes reform state, education, age, marriage age, household size, log of MPCE, urban dummy, caste, year of marriage dummies, and state dummies as control variables.

Standard errors in parentheses

*** p<0.01, ** p<0.05, * p<0.1

Source: Authors' calculation using IHDS - I (2004-05)

**Table A.5: Regression estimates of impact of HSAA on women's autonomy indicators for Rural Landowners**

| Dependent Variable | (1) Household | (2) Marriage Choice | (3) Physical | (4) Civil | (5) Economic | (6) IPV |
|---|---|---|---|---|---|---|
| **HSAA*Hindu** | - | 0.545*** | 2.533*** | 0.102 | 0.0814 | -0.584 |
| | - | (0.134) | (0.150) | (0.160) | (0.135) | (0.372) |
| **HSAA** | -0.903*** | -0.613*** | -1.813*** | 0.0155 | -0.182 | 0.556 |
| | (0.228) | (0.134) | (0.138) | (0.161) | (0.135) | (0.371) |
| **Hindu** | 0.341** | -0.118** | 0.486*** | -0.0561 | 0.448*** | -0.205* |
| | (0.145) | (0.0503) | (0.0491) | (0.0846) | (0.0462) | (0.113) |
| **Observations** | 34,491 | 37,744 | 37,761 | 37,706 | 37,769 | 37,341 |

Note: All regression includes reform state, education, age, marriage age, household size, log of MPCE, urban dummy, caste, year of marriage dummies, and state dummies as control variables.
Standard errors in parentheses
*** $p<0.01$, ** $p<0.05$, * $p<0.1$
Source: Authors' calculation using IHDS - I (2004-05)